\documentclass[preprint]{iucr}              
\usepackage{amssymb}
\usepackage{graphicx}

     \journalcode{M}              

\begin{document}                  



\title{Electronic Materials with Wide Band Gap: Recent Developments}
\shorttitle{Electronic Materials}


\cauthor{Detlef}{Klimm}{detlef.klimm@ikz-berlin.de}{address if different from \aff}

\aff{Leibniz Institute for Crystal Growth, Max-Born-Str. 2, 12489 Berlin \country{Germany}}









\maketitle                        

\begin{synopsis}
Usually semiconductors with band gap $E_g\approx3$\,eV or larger are called wide band gap materials. Then optical emission can span the whole visible spectral range, enabling the development of devices for solid state lighting. Besides, large the $E_g$ results in a high electrical breakthrough field which is interesting for high power electronics.
\end{synopsis}

\begin{abstract}
The development of semiconductor electronics is shortly reviewed, beginning with the development of germanium devices (band gap $E_g=0.66$\,eV) after world war II. Quickly a tendency to alternative materials with wider band gap became apparent, starting with silicon ($E_g=1.12$\,eV). This improved the signal/noise ratio for classical electronic applications. Both semiconductors have tetrahedral coordination, and by isoelectronic alternative replacement of Ge or Si with carbon or several anions and cations other semiconductors with wider $E_g$ are obtained, that are transparent for visible light and belong to the group of wide band gap semiconductors. Nowadays some nitrides, especially GaN and AlN, are the most important materials for optical emission in the ultraviolet and blue spectral region. Oxide crystals, such as ZnO and $\beta$-Ga$_2$O$_3$, offer similarly good electronic properties but suffer still from significant difficulties in obtaining stable and technically sufficient $p$-type conductivity.
\end{abstract}


\section{Introduction}

Semiconductors are crystalline or amorphous substances with a full valence band and an empty conduction band. Both bands are separated by the band gap energy $E_g$. Electronic charge transport in such systems is possible by fulfilling the following conditions:

\begin{enumerate}
\item Electrons must be emitted from the valence to the conduction band e.g. by thermal emission. ``Intrinsic conduction'' results then from the movement of these negative free electrons and the corresponding positive ``defect electrons'' (or holes) in opposite directions, if an electric field is applied. The conduction rises with temperature $T$ and becomes remarkable if the average thermal energy of electrons $k_\mathrm{B}T$ ($\approx$25\,meV at room temperature) approaches $E_g/2$. At sufficiently high $T$, this condition is fulfilled by every material.
\item Small amounts of suitable impurities can create additional ``dopant levels'' in the otherwise empty band gap. ``Shallow acceptor'' levels are situated close to the bottom of the gap, typically a few 10\,meV above the valence band. Consequently at room temperature they are filled almost completely by thermal emission, leaving behind holes in the valence band ($p$-type conductivity). For ``shallow donors'' close to the top of the gap the situation is opposite --- these levels can emit electrons into the conduction band ($n$-type conductivity). Deep acceptors or deep donors, which are situated close to the middle of the energy gap, do not contribute significantly to the electric carrier concentration, and thus to the electric conductivity.
\end{enumerate}

Insulators are materials with very large $E_g$, typically in excess of 3\ldots5\,eV. However, this limit is quite arbitrary and turns out to be subject of technical development: substances such as aluminum nitride or even diamond are nowadays usually considered to be semiconductors. Semiconductors with $E_g$ considerably larger than the ``normal'' semiconductors Si, Ge, or GaAs (see Tables \ref{tab:diamond} and \ref{tab:sphalerite}) are called wide band gap semiconductors, and are the topic of this article. In contrast, narrow band gap semiconductors have small $E_g$ of a few 100\,meV.

\section{A short look into history: germanium and silicon}

Faraday revealed already in the 1830s that some substances improve electric conductivity with $T$, which is in contrast to metals. However it took more than a century until semiconducting Ge crystals were grown with the Czochralski method \cite{teal50,teal51,uecker14} and paved the way to broad technical relevance of semiconductors. The growth of germanium bulk crystals with high crystalline perfection was a breakthrough, and the origin of today's semiconductor-based electronic industry. Unfortunately, $E_g$ is comparably narrow for germanium, leading to considerably large intrinsic conduction which cannot be controlled by $p-n$ junctions, and is the origin of electronic noise.

\begin{table}
\label{tab:diamond}
\caption{Semiconductors crystals with diamond structure}
\begin{tabular}{lllll}      
                      & DIAMOND        & SILICON & GERMANIUM & GREY TIN    \\
\hline
$a_0$ (nm)            & 0.3567         & 0.5431  & 0.5658    & 0.6489 \\
$T$-range ($^\circ$C) & $\lesssim1500$ & $<1414$ & $<938$    & $<13$ \\
$E_g$ (eV)            & 5.48           & 1.12    & 0.66      & 0.08 \\
$\lambda_g$ ($\mu$m)  & 0.226          & 1.11    & 1.87      & $>15$\\
\end{tabular}
\end{table}

Table~\ref{tab:diamond} summarizes data from \cite{kasap07,factsage64,glusker94} for the 4th main group elements crystallizing in the diamond structure, like germanium. There $a_0$ is the lattice constant. $T$-range means the limit where disintegration of the diamond phase occurs, which is for Si and Ge by melting, for C by transformation from metastable diamond to the thermodynamically stable graphite, and grey $\alpha$-Sn transforms to tetragonal $\beta$-Sn which is stable under ambient conditions. For every energy gap some $\lambda_g=hc/E_g$ can be calculated: the minimum optical wavelength for which the material is transparent. Compared to Ge, Si has a much broader $E_g$, which reduces intrinsic conduction and electronic noise of devices. Indeed, silicon (mainly Czochralski grown) is today the most important semiconductor material. Grey tin is technically not relevant, but diamond receives increasing attention as a truly wide band gap semiconductor. Poly- and single crystalline diamond is grown typically by chemical vapor deposition (CVD) processes, and devices such as diodes and transistors show excellent breakthrough stability up to as much as 20\,MV/cm. It is expected to outperform other wide band gap semiconductors such as 4H-SiC and GaN at $300^{\,\circ}$C by more than one order of magnitude \cite{hiraiwa13,shiomi96}.

\begin{figure}
\caption{Stacking of layers A--B--C (from bottom to top) in the diamond structure. One atom of each layer is hatched for a better demonstration of the stacking sequence.}
\includegraphics[width=0.4\textwidth]{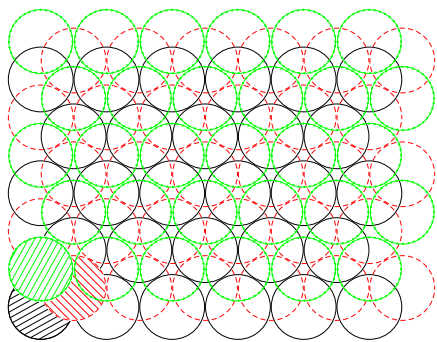}
\label{fig:stackings}
\end{figure}

The diamond structure is characterized by sp$^3$ hybrid orbitals which retract each other and are therefore directed from the central atom to the corners of a regular tetrahedron. The tetrahedra are arranged in layers, and if the position of the first layer (perpendicular to the $c$ axis) is designated as ``A'', subsequent layers are stacked in the somewhat shifted positions ``B'' and ``C'', resulting in a cubic stacking A---B---C---A---B---C (see Fig.~\ref{fig:stackings}). In contrast, the orbitals in the stable modification of carbon, graphite, are sp$^2$ hybridized. Here retraction directs the orbitals planar, with 120$^\circ$ angular distance. The remaining non-hybridized delocalized electron is situated out off the carbon plane and is the origin of almost metallic conductivity parallel to the basal (carbon atom) plane of its hexagonal structure. It is remarkable that carbon can form a large variety of other allotropes (graphene, nanotubes, buckminster-fullerenes, lonsdaleite) which are partially subject of intensive research, but did not yet reach remarkable technical relevance. In lonsdaleite carbon tetrahedra show hexagonal stacking A---B---A---B. In Fig.~\ref{fig:stackings} this means that the uppermost layer is again in the same position like the bottom layer.

\begin{figure}
\caption{The phase diagram silicon--carbon.}
\includegraphics[width=0.6\textwidth]{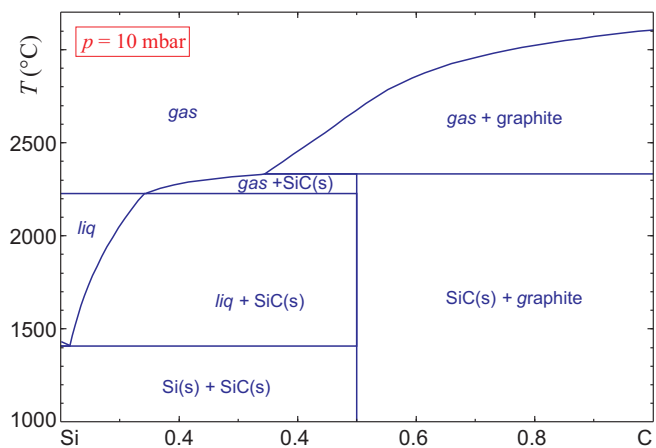}
\label{fig:PD-SiC}
\end{figure}

The diamond type elements of the fourth main group show partial (only in the case of Si--Ge complete) mutual solubility. Typically the solubility is larger (possibly under non-equilibrium conditions) for epitaxial layers. \cite{soref14} discussed the properties of alloys in the C--Si--Ge--Sn quaternary system and claimed that only alloys containing tin might offer direct band gaps. However, this publication disregarded the existence of the only intermediate compound, silicon carbide, in this system, which is an important semiconductor and will be discussed in the following section~\ref{sec:binary}. The homogeneity range of the three solid phases in the phase diagram Fig.~\ref{fig:PD-SiC} is only a few parts per million.

\begin{figure}
\caption{Derivation of tetrahedral multi-cation from element structures.}
\includegraphics[width=0.95\textwidth]{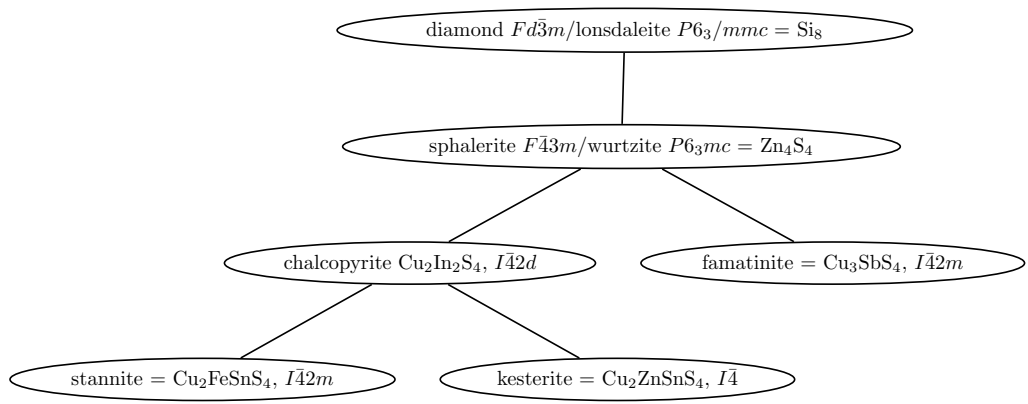}
\label{fig:tetrahedral}
\end{figure}

From the cubic diamond structure and the hexagonal lonsdaleite structure binary or ternary compounds respectively structures can be derived if the C-atoms are substituted in an ordered manner by other atoms in such a way that the average number of four electrons per atomic site is maintained. For an overview see e.g. \cite{parthe64,delgado98}. The structures of diamond, lonsdaleite, and derived tetrahedrally bound compounds are called adamantane types. Fig.~\ref{fig:tetrahedral} shows the interdependency of such tetrahedral structures with several sulfides as examples. It is obvious that the crystal symmetry drops with increasing chemical complexity. Only a few of these structure types, namely diamond, sphalerite, and wurtzite are found for wide-band gap semiconductors. Some others, such as kesterite and stannite with narrow $E_g$ around 1.0--1.5\,eV, are technically relevant e.g. as absorbers for thin-film solar cells \cite{redinger11}.

\section{Binary compound derived from diamond and lonsdaleite}
\label{sec:binary}

Such AB compounds comprise of alternating AB$_4$ (or A$_4$B, respectively) tetrahedra which are linked through their corners. Different stackings for the tetrahedron layers are observed, such as for diamond and lonsdaleite. If the diamond stacking is performed with AB$_4$ tetrahedra, the atom sites are identical with diamond itself, and just A- and B-atoms are alternating. The structure remains cubic, but symmetry is lowered to space group $F\bar{4}3m$. This is the sphalerite (= zincblende) structure type. In a similar way, lonsdaleite stacking of AB$_4$ tetrahedra results also in conservation of the atomic positions with alternating atom types. The resulting wurtzite structure belongs to space group $P6_3mc$. In an ideal wurtzite structure (stacking of ideal spheres) one has $c/a=\sqrt{8/3}=1.633$ which is not always fulfilled and results then in distorted tetrahedral bonding.

Often the type of stacking is fixed for one specific compound AB, because deviations from the ideal stacking increase the lattice energy of the crystal by the stacking-fault energy $\gamma$. Although usually $\gamma$ is given in energy per area units, scaling in energy per atom is preferred, as actually a certain number of bonds has to be broken to create the stacking fault. \cite{Gottschalk78} showed that $\gamma$ drops almost linearly if the ionicity of the A--B bonds rises and for cubic A$^\mathrm{III}$B$^\mathrm{V}$ compounds $\gamma$ ranges from $(53\pm7)$\,meV/atom for GaSb to $(17\pm3)$\,meV/atom for InP. Low $\gamma$ values are detrimental for crystal growth processes because already small thermal stresses can lead to stacking faults which impede electronic properties of the material.

The similarity of carbon and silicon is responsible for the low ionicity of SiC. Especially for this compound a huge variety of stacking orders can be observed, which are called polytypes. It turns out that different SiC polytypes are energetically almost identical, all of them (especially $\alpha$-SiC, see below) have very low stacking-fault energies \cite{hong00}. Consequently, they can easily coexist, can be transformed into each other, or switching between polytypes during growth occurs \cite{rost05}. Often the polytypes are described by the Ramsdell notation, which is a number giving the period of stacking followed by the letter H, C, or R which means that the stacking symmetry is hexagonal, cubic, or rhombohedral, respectively. Actually, silicon carbide can belong only to one of the four space groups $P3m1$, $R3m1$, $P6_3mc$, or $F\bar{4}3m$ \cite{krishna01}. Historically, the cubic (zincblende, 3C) silicon carbide is labeled $\beta$-SiC, whereas the other modifications are called $\alpha$-SiC. Table~\ref{tab:polytypes} reports some relevant SiC polytypes between the pure hexagonal 2H and pure cubic 3C extremes. The second line reports the average thickness of a single layer, which differs not much. All polytypes have rather large indirect band gaps, especially non-cubic $\alpha$-SiC. 

\begin{table}
\label{tab:polytypes}
\caption{Some polytypes of SiC \cite{bechstedt97,ching06,tairov83}}
\begin{tabular}{llllll}
                      & 2H (= wurtzite)  & 4H      & 15R      & 6H       & 3C (= sphalerite) \\
\hline
$a_0$ (nm)            & 0.3076           & 0.30817 & 0.30817  & 0.30817  & 0.43579 \\
$c_0/n$               & 2.524            & 2.5198  & 0.2520   & 0.2520   & 0.2517 \\
$E_g$ (eV)            & 3.33             & 3.27    & 2.986    & 3.02     &  2.39 \\
\end{tabular}
\end{table}

SiC is mechanically hard, chemically inert, and can be integrated well in standard semiconductor production lines. The growth of single crystals is a challenge, which can be seen already from the Si--C phase diagram in Fig.~\ref{fig:PD-SiC} that was calculated for a reduced pressure of $p=10$\,mbar. For significant larger or even ambient $p$ the ``\textit{liq}'' phase field for Si-rich compositions extends to higher $T$ which results then in peritectic melting of SiC to a Si-rich melt and solid carbon (graphite). Only under reduced $p\ll1$\,bar solid SiC is in equilibrium with ``\textit{gas}'' enabling sublimation growth (physical vapor transport, PVT) which is the standard growth technique for SiC single crystals. Besides, at $T\leq2300^{\,\circ}$C growth from melt solutions is an option (top seeded solution growth, TSSG) and was demonstrated and compared to PVT by \cite{hofmann99}. Often some metal (Fe, Ni, Cr, Ti, Li) is added to the melt. Different polytypes (e.g. 4H) are commercially available now as wafers with 150\,mm diameter, with $n$-type and $p$-type doping. The large band gap, good carrier mobility, and stability allow the production of electronic and optoelectronic devices with superior properties and for high breakdown field that are able to work even under harsh environment conditions. Besides it should be noted that SiC, under the name carborundum, is a mass product used e.g. as abrasive and for specialized ceramics in car brakes. Here, but also for electronics, its high thermal conductivity is beneficial which allows to remove lost heat.

Silicon carbide is the only tetrahedral bound semiconductor that can be derived from diamond or lonsdaleite by replacing C alternately with C or Si, respectively. If the structure is derived from diamond, one obtains the cubic zincblende (sphalerite) structure; if it is derived from lonsdaleite, the hexagonal wurtzite structure is obtained. It is remarkable that the names of both structure types are derived from zinc sulfide ZnS which can be found in nature in both structure types. Other isoelectronic replacements, with identical structural features, can be obtained by replacing C (4th group of the periodic system) alternately with elements from groups 3 and 5. Replacement with elements from main group 2 and 6 results mainly in compound with sodium chloride structure, with a few exceptions such as the insulator BeO \cite{austerman63} and the wide-band gap semiconductor MgTe \cite{kuhn71} belonging to the wurtzite type. Many subgroup elements, however, are also forming 2-valent ions --- the corresponding Me$^{2+}$ chalkogenides crystallize often in the sphalerite or wurtzite structure, and are semiconductors. Some of such A$^\mathrm{III}$B$^\mathrm{V}$ or A$^\mathrm{II}$B$^\mathrm{VI}$ semiconductors with technical relevance are shown in Table~\ref{tab:sphalerite}; even the A$^\mathrm{I}$B$^\mathrm{VII}$ compound silver iodide crystallizes below $\lesssim162^{\,\circ}$C in the wurtzite structure and has a wide band gap.

\begin{table}
\label{tab:sphalerite}
\caption{Semiconductors with sphalerite (S) or wurtzite (W) structure}
\begin{tabular}{llllllll}
                      & GaN       & GaP         & GaAs    & AlN     & ZnO            & ZnSe      & $\beta$-AgI  \\
\hline
type                  & W         & S           & S       & W       & W              & S         & W   \\
$a_0$ (nm)            & 0.319     & 0.5451      & 0.5653  & 0.311   & 0.3253         & 0.5668    & 0.458 \\
$c_0$ (nm)            & 0.519     & --          & --      & 0.498   & 0.5213         & --        & 0.7494 \\
$E_g$ (eV)            & 3.44      & 2.26        & 1.42    & 6.2     & 3.3            & 2.7       & 2.63 \\
$\lambda_g$ ($\mu$m)  & 0.36      & 0.59        & 0.87    & 0.20    & 0.38           & 0.46      & 0.47 \\
\end{tabular}
\end{table}

The compounds with larger ionicity tend to crystallize in the wurtzite structure, and the larger ionicity goes along with larger $E_g$. For GaP optical transparency reaches the visible spectral range, and wafers are transparent in the red. AlP (wider $E_g =2.45$\,eV) with sphalerite structure has relevance as semiconductors only in mixed crystals with other A$^\mathrm{III}$B$^\mathrm{V}$ compounds. Pure AlP, in contrast to other group 3 phosphides and arsenides tends to hydrolyze with moisture under formation of poisonous phosphine gas PH$_3$ and is used as pesticide. Gallium and indium phosphides, arsenides and partially antimonides for semiconductor applications are typically grown as bulk single crystals from the melt, either by crystallization inside a crucible from the bottom to the top (Bridgman, Vertical Gradient Freeze = VGF) or by pulling (Czochralski). Arsenides and the more phosphides tend to have large arsenic or phosphorous vapor pressure (up to several 10\,bar) at their melting points. Disintegration of the semiconductor compounds can be avoided by overpressure and ``liquid encapsulation'' of the material with B$_2$O$_3$ which melts at $450^{\,\circ}$C significantly below the semiconductor and forms a liquid layer on top of the melt.

\begin{figure}
\caption{Temperature--pressure phase diagram of AlN demonstrating decomposition AlN$\rightarrow$Al + $\frac{1}{2}$N$_2$ at insufficient pressure. Calculated with \cite{factsage64}.}
\includegraphics[width=0.6\textwidth]{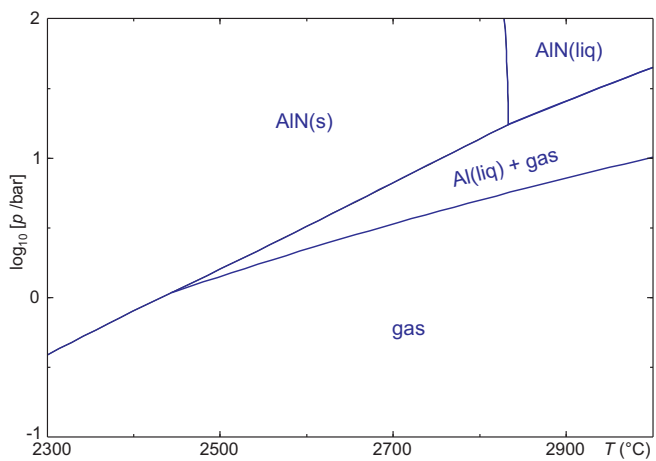}
\label{fig:Al-N2}
\end{figure}

Among the group 3 nitrides, BN has not yet reached its full potential. Different modifications occur, for the wurtzite type $E_g=5.2$\,eV is reported which makes the material almost an insulator. An excellent database on this interesting compound can be found on the web \cite{ioffe14}. For the other group 3 elements the affinity to nitrogen decreases in the order Al -- Ga -- In which results in decomposition of the nitrides upon heating below their melting points, and actually InN is so far only relevant as admixture to (Al,Ga,In)N mixed crystals because the growth of single crystals is difficult. InN layers were obtained by HVPE, a technique that will be explained below in the context of GaN \cite{sato94} and InN nanowires were obtained from the gas phase in a vapor-liquid-solid (VLS) process \cite{tang04}. The stability of AlN is shown in the $T-\log[p]$ phase diagram in Fig.~\ref{fig:Al-N2} where below atmospheric pressure AlN(s) in in equilibrium with gas only (sublimation), at intermediate pressure with gas (containing N$_2$ + Al) and remaining molten Al, and only at high $p$ melting of AlN does occur. The calculated triple point is here $2830^{\,\circ}$C and 17.4\,bar. The accurate position of this triple point is still under discussion, and some other references claim high $p_{\mathrm{N}_2}$ beyond 100\,bar \cite{ioffe14AlN}, but it should be acknowledged that it is almost impossible to measure exact values under such extreme conditions. In experiments AlN decomposition starts at significantly lower $T$ than the AlN(s) phase boundary in Fig.~\ref{fig:Al-N2} if the material is heated in other gases than N$_2$. The current author obtained already 10\% mass loss of a 33\,mg AlN sample that was heated in a DTA/TG apparatus in helium atmosphere to $2040^{\,\circ}$C (unpublished result).

The extreme conditions that are required to maintain solid/liquid equilibrium for AlN make sublimation growth more feasible, and indeed it is typically performed at $T>2040^{\,\circ}$C and $p\lesssim1$\,bar \cite{hartmann13}. For GaN the establishment of suitable growth conditions is more difficult, because gallium (in contrast to aluminum) does not evaporate sufficiently for sublimation growth. Actually Ga is the chemical element with the broadest range of liquid phase stability under ambient pressure: $29.8\leq T(^\circ\mathrm{C})\leq2203$ \cite{factsage64}. However, \cite{karpinski84} showed that at high $p$ and $T$ nitrogen does significantly dissolve in liquid gallium; 1\,mol\% N$_2$ solubility were found at $1500^{\,\circ}$C and 16\,kbar which proved sufficient for establishing melt solution growth of bulk GaN. Other technologies for the growth of bulk GaN rely on chemical transport of gallium species: From a supercritical ammonia solution phase 2-inch GaN crystals can be grown nowadays \cite{dwilinski10}.

\begin{figure}
\caption{Predominance diagram of the system Ga--N--H--Cl for prevailing NH$_3$ fugacity 1\,bar. Calculated with \cite{factsage64}.}
\includegraphics[width=0.6\textwidth]{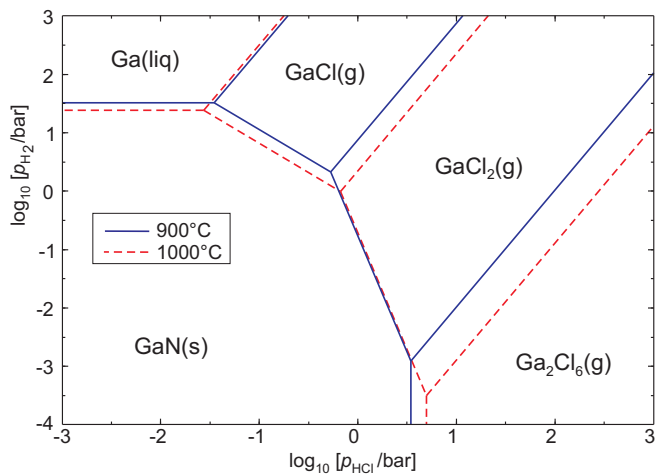}
\label{fig:HVPE}
\end{figure}

Hydride Vapor Phase Epitaxy (HVPE) is an epitaxy process for the deposition of semiconductor layers, including GaN. For this process metallic gallium reacts at ca. $850^{\,\circ}$C with a HCl flow under formation of gaseous gallium(I) chloride: Ga + HCl $\rightarrow$ GaCl + 0.5 H$_2$. After passing the Ga source, the GaCl/HCl flow (with N$_2$ as carrier gas) reacts with ammonia and gallium nitride is deposited: GaCl + NH$_3$ $\rightarrow$ GaN + HCl + H$_2$. Fig.~\ref{fig:HVPE} shows that under process conditions GaCl gas is in equilibrium with solid GaN and that the latter is formed if the HCl fugacity drops distant from the source. Equilibrium ($\it\Delta G= $ 0) for the GaN formation reaction given above is reached under ambient pressure at $918^{\,\circ}$C. Although HVPE is in principle a layer growth process, it is suitable for the production of bulk material too, with growth rates around 100\,$\mu$m/h and up to several millimeters thickness. However, some drawbacks have to be taken into account: 1) as a side reaction between HCl and NH$_3$ large quantities of solid NH$_4$Cl are formed that can obstruct the system; 2) HVPE grown GaN is bowed, which interferes the production of planar wafers \cite{lipski12}. \cite{jacobs10} showed that in systems containing graphite the halogen \{Cl\} can be replaced by the pseudo-halogen \{CN\} and gaseous gallium(I) cyanide GaCN transports Ga. Crystalline GaN is then deposited either by the thermal decomposition of GaCN or by a reaction that is analogous to that of the common HVPE growth technique given above.

\cite{strite92}, and more recently \cite{oleary06} with a deeper insight on electronic properties, reviewed AlN, GaN, InN and their solid solutions which are covering a wide range $6.2\geq E_g(\mathrm{eV})\geq0.68$. Meanwhile remarkable technical progress was reached, and now (Al,Ga,In)N based devices are the basis for solid state lighting applications. Because GaN and AlN substrates are still scarce and expensive, homoepitaxy plays no significant role and is still used mainly for basic research \cite{Funato12}. Heteroepitaxy is performed on different surfaces of $\alpha$-Al$_2$O$_3$ (sapphire), mainly $(0001)$ but also $(21\bar{3}1)$ and $(11\bar{2}0)$. Other useful substrates are several A$^\mathrm{III}$B$^\mathrm{V}$ compounds such as GaAs, SiC polytypes, and ZnO. LiAlO$_2$ and LiGaO$_2$ are interesting alternatives because epitaxial misfit is much smaller compared e.g. to sapphire and large bulk single crystals for substrates up to 2 inch diameter are available too. Besides, after epitaxy the substrates can easily be dissolved in diluted acids which makes contacting of epitaxial layers from both sides feasible \cite{liu04,velickov08}. Fascinating new possibilities are offered by the integration of GaN in silicon technologie, especially for High-Electron-Mobility Transistors (HEMTs) \cite{hu14}. Even if the lattice mismatch for (0001) GaN on (111) Si is as large as 17\%, satisfactory layers can be grown in such ``GaN-on-Si'' process by metal-organic chemical vapor deposition (MOCVD) using graded buffer layers \cite{drechsel12}.

Among the other substances in Table~\ref{tab:sphalerite} zinc oxide has by far the greatest practical impact nowadays. It is a typical direct wide band gap semiconductor and its properties are reviewed in numerous articles \cite{look07,janotti09,klimm11}. Such as for some other oxide semiconductors, the electronic properties of the surface is significantly different from the bulk, and can be manipulated by doping or adsorbance layers. The latter effect is used for gas sensing applications, whereas ZnO ceramics with small additives of Bi$_2$O$_3$ and other oxides have an extremely nonlinear resistance resulting from the grain/interlayer/grain boundaries. This nonlinearity is so large that ceramic ``varistors'' are commercially produced with negligible resistance below and almost infinite resistance above some threshold voltage.

Like most other oxide semiconductors, ZnO is intrinsic $n$-type. Numerous attempts to obtain stable $p$-type conductivity with technically sufficient hole concentration and mobility failed so far --- which is a severe drawback compared to Al-Ga-In nitride and restricts or even prohibits manufacturing of many devices. If the $n$-type conductivity of ZnO is enhanced e.g. by doping with aluminum, transparent electrodes e.g. for solar cell applications or flat screen panels can be produced which are much cheaper compared to ITO (indium-tin-oxide) electrodes \cite{kluth99}. \cite{grundmann10} reports an electron concentration around $10^{21}$\,cm$^{-3}$, Hall mobility 47.6\,cm$^2$/Vs and a resulting specific resistivity of $8.5\times10^{-5}$\,$\Omega$cm.

\section{Other oxides}

ZnO is the only semiconducting oxide material which can be found in Table~\ref{tab:sphalerite}. As a result of the large electronegativity of oxygen (3.5) compared to the anions of classical semiconductors (sulfur: 2.5; phosphorous: 2.1; arsenic: 2.0) oxides tend to have comparably high ionicity and wide $E_g$. Nitride semiconductors (electronegativity of N: 3.0) are in this respect intermediate between oxides and classical semiconductors.

Many metal oxides are true isolators with large band gap, such as $\alpha$-Al$_2$O$_3$ (corundum). The elements that are following after aluminum in group III of the periodic system, and some other elements such as tin, lead, bismuth, titanium, have oxides with $E_g$ values that fall into the range of wide band gap semiconductors. The terms ``Transparent Conducting Oxide'' (TCO) or ``Transparent Semiconducting Oxide'' (TSO) are often used for such substances which are combining optical transparency with electrical transport properties.

\begin{table}
\label{tab:oxides}
\caption{Some wide band gap oxides}
\begin{tabular}{llllll}
                      & $\alpha$-Al$_2$O$_3$  &  $\beta$-Ga$_2$O$_3$ & In$_2$O$_3$  & SnO$_2$    & CuAlO$_2$  \\
\hline
structure             & corundum              & monoclinic           & bixbyite     & rutile     & delafossite   \\
space group           & $R\bar{3}c$           & $C2/m$               &  $Ia\bar{3}$ & $P4_2/mnm$ & $R\bar{3}m$   \\
$a_0$ (nm)            & 0.51284               & 1.2214               & 1.0117       & 0.47397    & 0.2857     \\
$b_0$ (nm)            & --                    & 0.30371              & --           & --         & 0.2857     \\
$c_0$ (nm)            & --                    & 0.57981              & --           & 0.31877    & 1.6939     \\
angle ($^\circ$)      & $\alpha=55.28$        & $\beta=103.83$       & --           & --         & --        \\
$E_g$ (eV)            & 8.3                   & 4.8                  & 3.6          & 3.6        & 2.22    \\
\end{tabular}
\end{table}

An increasing number of TCO and TSO compounds were studied during the last decade, some of them can be found in Table~\ref{tab:oxides}. \cite{ramesh08} reviewed status and prospects of ``Oxide Electronics'', which brought already some remarkable results with technical relevance. The replacement of SiO$_2$ for MOSFET gates by ``high-$\kappa$'' materials enabling larger packing density of circuits is an instructive example: initially experiments focused on well known substances such as BaTiO$_3$, but failed because the gates were degrading. \cite{hubbard96,schlom02} performed a thermodynamic search  for metal oxides that are stable in contact with silicon, which has a high oxygen affinity. About one decade after their invention that hafnium oxide HfO$_2$ belongs to the few compounds which might replace SiO$_2$, this high-$\kappa$ material was introduced into the production of devices. Thermodynamic equilibria have to be considered not only for the implementation of oxides into silicon electronics. Also for the bulk or layer growth of oxides redox equilibria play a major role --- more than typically for other anions such as nitride or sulfide. 

This difference can be explained mainly by the vast number of different metal oxides that exist especially for the subgroup elements, with consequently smaller phase stability fields for each of them. Among the 4776 compounds of the FactPS thermodynamic database \cite{factsage64} can be found e.g. for manganese: 4 oxides, 2 sulfides, 2 nitrides; for copper: 2 oxides, 2 sulfides, 1 nitride; for titanium: 12 oxides, 5 sulfides, 1 nitride.

\begin{figure}
\caption{Predominance diagram of the system Cd--O--S for two temperatures.}
\includegraphics[width=0.95\textwidth]{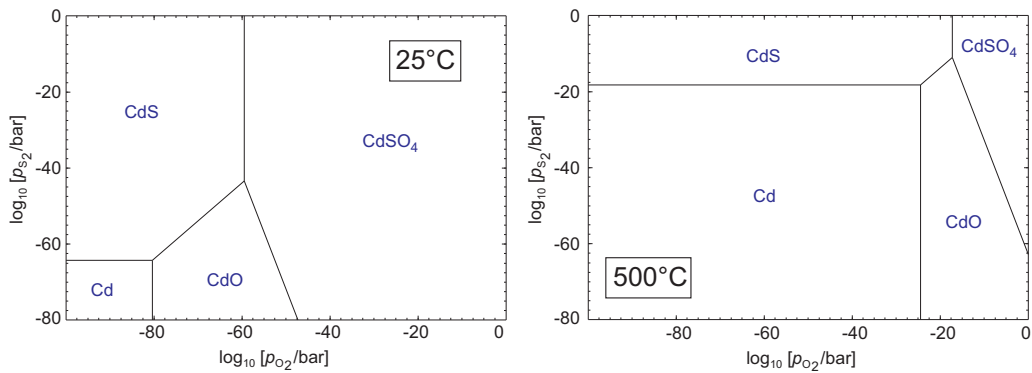}
\label{fig:Cd-O-S}
\end{figure}

If metal oxides with oxidation states $m$ and $m+1$ can transform via the redox equilibrium
\[
2\,\mathrm{MeO}_{m/2} + \frac{1}{2}\,\mathrm{O}_2 \rightleftarrows 2\,\mathrm{MeO}_{(m+1)/2}
\]
then the Gibbs free energy change of this reaction is proportional to $\it{\Delta}G = RT\ln[p_{\mathrm{O}_2}]$ if the fugacity of both oxides can be neglected. Plots $RT\ln[p_{\mathrm{O}_2}](T)$ are linear and separate predominance fields for the subsequent metal oxides \cite{Pelton91,Klimm09}. In a similar manner the behavior of one or more metals in dependence of several nonmetal fugacities can be calculated and results in predominance diagrams ($T=$\,const.) with straight phase boundaries. Fig.~\ref{fig:HVPE} explained by such diagram the HVPE process for GaN, and Fig.~\ref{fig:Cd-O-S} shows equilibria between cadmium, oxygen and sulfur for two different temperatures.

In modern epitaxial systems heterostructures of almost every composition can be grown and under sufficiently low $T$ also nonequilibrium states can be produced because they are metastable. One should be aware, however, that with time and especially if $T$ increases e.g. in an active device, metastable structures might approach equilibrium. This exactly happened with BaTiO$_3$ MOSFET gates before the introduction of HfO$_2$! From Fig.~\ref{fig:Cd-O-S} one reads that CdO and CdS are in equilibrium for a wide $T$ range and consequently heterostructures of oxide and sulfide are possible. This agrees with experimental results \cite{Li09}. On the other side, for very high fugacities of O$_2$ and S$_2$ the sulfate CdSO$_4$ is a stable intermediate phase between oxide and sulfide. Such optional intermediate phase should not be forgotten for other heteroepitaxial systems such as nitride an oxide \cite{shimamura05}. The observation of NO$_x$ bond signatures by photoelectron spectroscopy of oxidized InN surfaces \cite{eisenhardt12} could hint on the formation of indium nitrate In(NO$_3$)$_3$ or nitrite In(NO$_2$)$_3$ as intermediate phase between InN and In$_2$O$_3$.

Chemical stability considerations play a major role also for the bulk growth of oxide crystals that can be used for substrates. At least ZnO, Ga$_2$O$_3$, In$_2$O$_3$ and SnO$_2$ have in common that a high (in the case of SnO$_2$ even not accessible) melting point exceeding $1800^{\,\circ}$C is combined with a comparably high $p_{\mathrm{O}_2}$ that is necessary to avoid decomposition of the MeO$_x$ oxide to metal Me and oxygen. If formed, the free metal would immediately alloy to iridium which is the standard crucible material for oxide crystal growth in this $T$ range. Because the relevant Me--Ir systems contain low melting eutectics, this alloying destroys the crucible.

Certainly for ZnO the broadest variety of methods was published that allow to circumvent the stability problem mentioned above, and these methods are more or less suitable also for other wide band gap semiconducting oxides. These methods include:
\begin{itemize}
\item Growth from solutions, either water or ammonia based (often hydrothermal, ammonothermal conditions) --- or from molten salts (e.g. Top Seeded Solution Growth, TSSG).
\item Growth by Physical Vapor Transport (PVT, sublimation) or Chemical Vapor Transport (CVT)
\item Growth from the melt either from cold crucibles (``skull melting'') or from hot iridium crucibles with a ``reactive atmosphere''.
\end{itemize}
and where reviewed elsewhere \cite{klimm11}.

Crystal growth methods from the melt, such as Czochralski or Bridgman, combine chemical purity (no solvents used) with high growth rate. Indeed, bulk growth of the most important semiconductors silicon and gallium arsenide is exclusively performed from the melt. Meanwhile also ZnO \cite{schulz06}, $\beta$-Ga$_2$O$_3$ \cite{villora04,aida08,galazka10}, and In$_2$O$_3$ \cite{galazka13b} where grown this way. Tin(IV) oxide SnO$_2$ could not be melt-grown so far: The melting point of this substance is for sure much higher than around the $1630^{\,\circ}$C, that are given in several references and databases, cf. \cite{factsage64}. If heated up to this $T$-range, the edges of SnO$_2$ powder grains become rounded, and in DTA measurements one even can see an endothermal effect that might be mixed up with melting. Studying the phenomena in more detail shows, however, that this claim is wrong and locally different partial evaporation of the material takes place instead. Also heating in pressurized chambers ($\lesssim20$\,bar) beyond $2000^{\,\circ}$C did not result in molten SnO$_2$ ingots. 

\begin{figure}
\caption{Minimum oxygen fugacity of several transparent conducting oxides at their melting points $T_\mathrm{f}$ compared to the $p_{\mathrm{O}_2}(T)$ that results from the thermolysis of carbon dioxide.}
\includegraphics[width=0.6\textwidth]{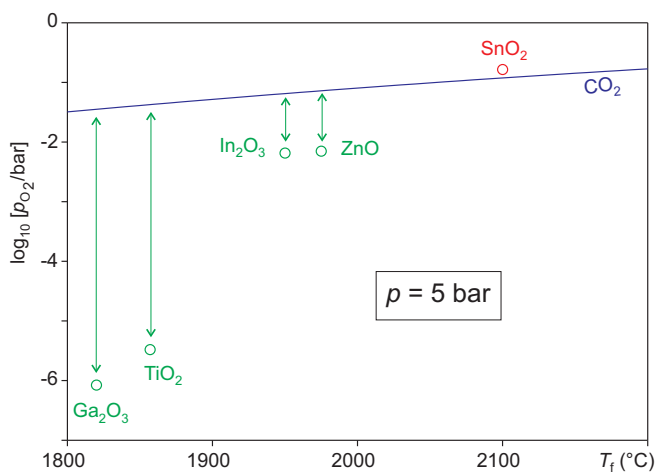}
\label{fig:oxides}
\end{figure}

\cite{Klimm09} reported that the partial thermal dissociation of carbon dioxide CO$_2$ $\rightleftarrows$ CO + 0.5\,O$_2$ gives the basis for melt crystal growth of oxides in a reactive atmosphere. The reaction equilibrium shifts to the product side if the temperature is increased. Consequently pure CO$_2$ or (for somewhat lower oxygen fugacity) CO$_2$/CO mixtures can give such a ``self-adjusting atmosphere'' where $p_{\mathrm{O}_2}(T)$ meets for all $T$ the stability field of the desired oxide. For the semiconducting oxides mentioned so far, comparable large $p_{\mathrm{O}_2}(T)$ are required, and Fig.~\ref{fig:oxides} explains why bulk crystal growth of SnO$_2$ from the melt is so difficult: The required oxygen fugacity is so high, that it cannot be produced by CO$_2$ dissociation. Such as reported by \cite{galazka14}, not the high melting point itself is the problem, but oxide phase instability becomes an issue if the required oxygen fugacity approaches the $p_{\mathrm{O}_2}(T)$ line of CO$_2$, and growth is probably impossible this way if it is beyond the line.

For $\beta$-Ga$_2$O$_3$ large bulk crystals (EFG grown ribbons, float zone and Czochralski boules) are available and epitaxial techniques (MOCVD, MBE) are developed. \cite{higashiwaki12} reported the production of a metal-semiconductor field-effect transistor (MESFET) out of this material, and a high on/off drain current ratio $\approx10000$ was reached. The authors claimed that very high breakdown fields around 8\,MV/cm should be feasible with $\beta$-Ga$_2$O$_3$, which is almost as good as diamond (10--20\,MV/cm) and outperforms the current high-power material 4H-SiC (2.5\,MV/cm) and also GaN (3.3\,MV/cm). Monoclinic $\beta$-Ga$_2$O$_3$ is the stable modification between the melting point and room temperature, enabling melt growth of bulk crystals. With epitaxial growth on sapphire ($\alpha$-Al$_2$O$_3$) substrates and by alloying, $\alpha$-(Al,Ga,In)$_2$O$_3$ layers can be grown which allow band gap tuning from 3.8\,eV to 8.8\,eV \cite{fujita14}.

Nearly all oxide semiconductors are intrinsic $n$-type, which is a result of the strong localization of holes (if formed by doping or nonstoichiometry) at the oxide ions and impedimental for the development of devices with $p-n$ junctions. This is a general problem for all oxide semiconductors which can hardly be overcome completely, but some circumstances such as tetrahedral coordination of oxide ions and some degree of covalency can improve $p$-type conductivity \cite{banerjee05}. \cite{kawazowe97} demonstrated that delafossite type CuAlO$_2$ combines encouraging $p$-type conductivity with transparency in visible light. The carrier density of $1.3\times10^{17}$\,cm$^{-3}$ and Hall mobility for holes of 10.4\,cm$^2$\,V$^{-1}$\,s$^{-1}$ was explained by large hybridization of oxygen $2p$ orbitals with the $3d^{10}$ electrons of the Cu$^+$ closed shell, leading finally to low hole effective mass. These ideas were the basis of an extended numerical study by \cite{hautier13}. For 3052 binary and ternary oxides, either minerals or synthesized materials, DFT computations with the Vienna software package were performed to identify substances with low hole effective mass and large band gap. As expected, low hole masses are much scarce than low electron masses, but some substances which are so far rarely studied could be promising: K$_2$Sn$_2$O$_3$, Ca$_4$P$_2$O, Tl$_4$V$_2$O$_7$, PbTiO$_3$, ZrOS, B$_6$O, and Sb$_4$Cl$_2$O$_5$. It should be noted that the well known good hole mobility of Cu$_2$O (which has, however, small $E_g\approx2.1$\,eV) and CuAlO$_2$ could be reproduced too.

\begin{figure}
\caption{Ellingham predominance diagram of the system Cu--Al--O$_2$ with [Cu]:[Al] = 1:1.}
\includegraphics[width=0.6\textwidth]{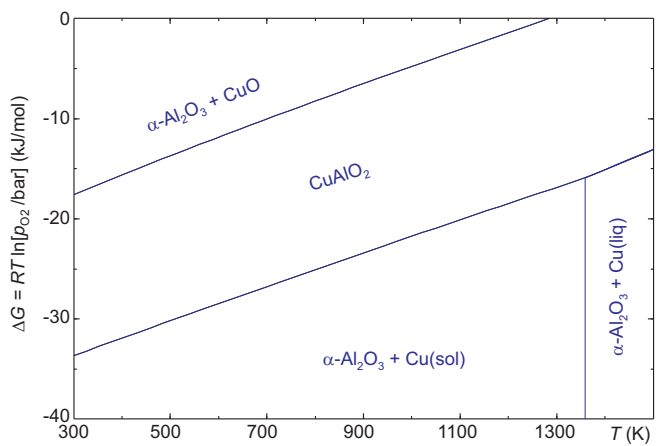}
\label{fig:Cu-Al-O2}
\end{figure}

Copper exists in most compounds as Cu$^{2+}$, and the untypical low valency Cu$^+$ is directly related to the $p$-type conductivity of CuAlO$_2$ and Cu$_2$O. In a similar manner, SnO$_2$ is the ``normal'' tin oxide and $n$-conducting, whereas potassium stannate(II) K$_2$Sn$_2$O$_3$ \cite{hautier13} such as tin(II) oxide SnO shows hole conductivity \cite{ogo08}. It is certainly possible to obtain small quantities of such ``low valency'' oxides, either as epitaxial layers or in the bulk, just by crystallizing them together in an oxygen poor atmosphere. This was demonstrated by \cite{yoon13} with millimeter-sized CuAlO$_2$ crystal that could be grown from an Cu$_2$O melt flux excess. Here the material was kept several days at the growth temperature beyond $1100^{\,\circ}$C and then cooled obviously relatively quickly to room temperature. Already \cite{gadalla64} showed that in air CuAlO$_2$ becomes unstable below $1030^{\,\circ}$C under partial oxidation to CuO. The stability diagram in Fig.~\ref{fig:Cu-Al-O2} was calculated with \cite{factsage64}. It demonstrates that the Cu(I) compound CuAlO$_2$ has a stability field between metallic copper for lower, and CuO for higher $p_{\mathrm{O}_2}$. The calculation of such diagrams requires that thermodynamic data such as $G(T)$ are available for the intermediate phase, here CuAlO$_2$. But even if this is not the case, a coarse approximation could be reached if the relevant phase is simply neglected for the equilibrium calculation. Then an intermediate phase field ``$\alpha$-Al$_2$O$_3$ + Cu$_2$O'' appears instead at almost the same place --- just the upper and lower phase boundaries are shifted ca. 5\,kJ/mol inwards. This is because the formation energy Al$_2$O$_3$ + Cu$_2$O $\rightarrow$ 2\,CuAlO$_2$ is not taken into account. The major contributions to the Gibbs free energy of the system, however, result from the equilibria between the subsequent oxidation states of copper. Hence it is almost sufficient to have $G(T)$ date for all relevant element oxides available.

It should be noted that the reliability of this diagram is expected to be satisfactory especially for high oxygen fugacity, because the underlying measurements were performed by \cite{gadalla64} in air or even more oxidizing atmosphere and more recent results are not published. Consequently data below $\approx20$\,kJ/mol are extrapolated. Unpublished measurements of the current author indicate, however, that this interpolation is not far from reality. One can expect that working in a suitable reactive atmosphere will pave the way to bulk crystal growth conditions where the delafossite phase can be kept thermodynamically stable.

\section{Summary and conclusion}

The successful story of technical semiconductor applications started in the early 1950s with the first $p-n$ junctions that were made inside Czochralski grown germanium single crystals. For electronic applications germanium is nowadays replaced almost completely by silicon. Optoelectronics, mainly based on A$^\mathrm{III}$B$^\mathrm{V}$ compound such as GaAs, opened in the 1970s a new field for semiconductors, but arsenides and phosphides which could be grown in that time have a narrow band gap enabling optical emission only from the infrared to the green spectral range. This is sufficient e.g. for displays, indicators, and optical data transmission, but not for general illumination as the blue spectral range is missing.

The successful growth of several nitrides beginning in the late 1980s, especially the wide band gap semiconductors GaN and AlN on sapphire substrates, widened the accessible wavelength range into the ultraviolet region. Moreover, white light can now be produced for ``solid state lighting'', with positive impact on global energy consumption by replacing incandescent light bulbs with light emitting diodes.

The substance palette is extended by different polytypes of silicon carbide, SiC, and several oxides, such as ZnO, $\beta$-Ga$_2$O$_3$, and In$_2$O$_3$. These wide band gap semiconductors are used as substrates for layer deposition of the more classical semiconductors mentioned above, as well as for active devices. Unfortunately, the still unsatisfactory $p$-type conductivity of semiconducting oxides is an issue which hinders the development of devices significantly. Wide band gap semiconductors such as SiC, GaN, and $\beta$-Ga$_2$O$_3$ have potential not only for optoelectronics, but also for high power devices.

For the recently reported organic-anorganic substances such as CH$_3$NH$_3$PbBr$_3$ \cite{kojima09} and CH$_3$NH$_3$SnI$_3$ \cite{hao14} with perovskite structure band gap tuning is possible by substitution of the halide and/or metal ion. They were used as absorbers in solar cell structures, enabling power conversion efficiencies greater than 15\%. We can hope that further progress is possible here, but it seems too early to include this substance group into this review.





%
%
%



\ack{Acknowledgements}

The author expresses his gratitude to M. Bickermann for reading the manuscript and for helpful comments. Z. Galazka is thanked for inspiring discussions on the fascinating field of transparent conducting oxides.







\end{document}